\documentclass[pra,aps,reprint,showpacs,superscriptaddress]{revtex4-1}

\usepackage{graphicx}
\usepackage{amsmath}

\usepackage[usenames]{color}
\usepackage{ulem}

\def\aop{\hat{a}}

\def\bop{\hat{b}}
\def\Dop{\tilde{D}}

\def\Spmop{\tilde{S}_\pm}

\def\Apmop{\tilde{A}_\pm}

\def\sigop{\hat{\sigma}^-}

\newcommand{\braket}[1]{\left\langle #1 \right\rangle}
\newcommand{\gamp}{\gamma_\perp}
\newcommand{\Dela}{\Delta_{\rm a}}

\newcommand{\lsz}{\left[}
\newcommand{\rsz}{\right]}
\newcommand{\lk}{\left(}
\newcommand{\rk}{\right)}

\newcommand{\gamll}{\gamma_\parallel}

\newcommand{\Eb}{{\cal E}_b}

\begin{document}
\title{Cavity dark mode of distant coupled atom-cavity systems}
\author{Donald H. White}
\affiliation{Department of Applied Physics, Waseda University, 3-4-1 Okubo, Shinjuku, Tokyo 169-8555, Japan}
\author{Shinya Kato}
\affiliation{Department of Applied Physics, Waseda University, 3-4-1 Okubo, Shinjuku, Tokyo 169-8555, Japan}
\affiliation{JST PRESTO, 4-1-8 Honcho, Kawaguchi, Saitama 332-0012, Japan}
\author{Nikolett N\'{e}met}
\affiliation{Dodd-Walls Centre for Photonic and Quantum Technologies, New Zealand}
\affiliation{Department of Physics, University of Auckland, Auckland 1010, New Zealand}
\author{Scott Parkins}
\affiliation{Dodd-Walls Centre for Photonic and Quantum Technologies, New Zealand}
\affiliation{Department of Physics, University of Auckland, Auckland 1010, New Zealand}
\author{Takao Aoki}
\email{takao@waseda.jp}
\affiliation{Department of Applied Physics, Waseda University, 3-4-1 Okubo, Shinjuku, Tokyo 169-8555, Japan}

\begin{abstract}
We report on a combined experimental and theoretical investigation into the normal modes of an all-fiber coupled cavity-quantum-electrodynamics system. The interaction between atomic ensembles and photons in the same cavities, and that between the photons in these cavities and the photons in the fiber connecting these cavities, generates five non-degenerate normal modes. We demonstrate our ability to excite each normal mode individually. We study particularly the `cavity dark mode', in which the two cavities coupled directly to the atoms do not exhibit photonic excitation. Through the observation of this mode, we demonstrate remote excitation and nonlocal saturation of atoms.

\end{abstract}
\maketitle

A future quantum internet depends on the connection and entanglement of many distant qubits~\cite{Kimble2008,Reiserer2015}. These qubits form the nodes of the network, and communication between nodes is carried via channels which transmit quantum information.
When coupling between the nodes is bi-directional~\cite{Serafini2006}, instead of uni-directional~\cite{Cirac1997, Ritter2012},
the system oscillates as a collective whole, and the oscillations can be projected onto a set of orthogonal normal modes in which energy is continuously exchanged between oscillators. This normal mode behavior defines the structural basis of the system dynamics, underlying the higher-level dynamical effects leading to {\it e.g.} operation of quantum gates~\cite{Cao2018} and the physical implementation of systems of strongly interacting photons~\cite{Torma1998, Hartmann2006, Greentree2006}.

All-fiber atom--cavity quantum electrodynamics (QED) systems, in which atoms are coupled to the cavity field via evanescent coupling through a tapered optical nanofiber region, are an especially attractive prospect for quantum networking due to the ease of connecting many nodes together in any arbitrary network configuration with minimal loss. A cavity QED system is typically formed by coupling the atoms to an in-fiber cavity formed by either two Fiber Bragg Gratings (FBGs)~\cite{LeKien2009,Wuttke2012,Yalla2014,Kato2015,Li2017,Keloth2017}, or else a ring cavity coupled via a fiber beamsplitter~\cite{Jones2016,Ruddell17,Schneeweiss2017}. This paper is focused on `coupled-cavities quantum electrodynamics', concerning the interaction of atoms coupled via cavity fields. Specifically, we focus on the properties of the normal modes of two atomic ensembles coupled via three optical cavities.

A dark mode is a class of normal modes in which one or more oscillators does not exhibit excitation due to destructive interference. An example of such a mode is a dark atomic state, which is prevented from absorbing a photon due to coupling induced by control fields~\cite{Lukin2003}. In addition to the widely used application of electromagnetically-induced transparency~\cite{Boller1991}, the dark mode of a coupled system has for example been used to suppress mechanical dissipation in an optomechanical resonator~\cite{Dong2012}. We recently demonstrated the `fiber dark mode' of a coupled-cavity QED system, where distant atoms interact with delocalized photons~\cite{Kato2018}. We show in this paper that another type of dark normal mode exists in this system, in which the photonic excitations at the atom locations are dark, such that the atoms are not locally exposed to light fields. This `cavity dark mode' is robust and does not depend on cavity symmetry. With the absence of local photons, we demonstrate nonlocal excitation and saturation of atoms.

\begin{figure}[t]
\centering
\includegraphics[width=8.2cm]{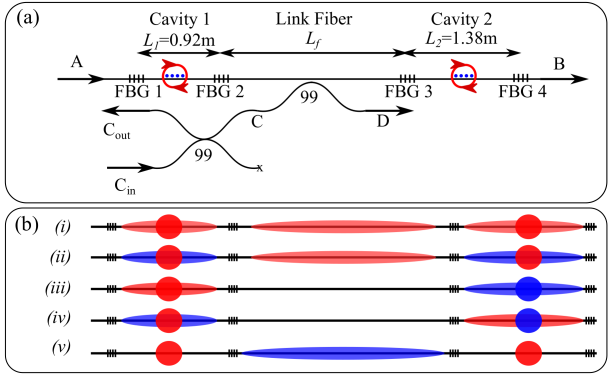}
\caption{(a) Schematic of the setup. Three optical cavities, comprising four Fiber Bragg Grating (FBG) mirrors, are connected in series. Optical nanofiber regions are fabricated within the two end cavities, enabling coupling to ensembles of atoms through the evanescent field. The system may be probed from either the cavity end ($A$) or from the central fiber beamsplitter ($C$), while the excitation is simultaneously detected at ports $B$ and $C$. (b) Schematic of normal modes of the system. Ellipses indicate cavity excitations, and circles indicate atom excitations. Red and blue are $\pi$ out of phase. Five normal modes are present: (i,ii) are symmetric bright modes, (iii,iv) are fiber dark modes, and (v) is the cavity dark mode.}
\label{fig:setup}
\end{figure}

The experiment comprises an elementary all-fiber quantum network, similar to our previous setup~\cite{Kato2018}, in which two nanofiber cavity QED systems are connected by an intermediate link fiber cavity, as illustrated in Fig.~\ref{fig:setup}(a) (in this paper, the two cavities directly coupled to atoms are named `cavities', while the linking fiber cavity is referred to as the `fiber'). Optical cavities are formed within the single-mode optical fiber between FBG mirrors, and atoms are coupled to the cavities via tapered fiber regions of diameter 400 nm. We experimentally excite and detect the five normal modes of this network of five coupled oscillators (three optical cavities and two $^{133}$Cs atomic ensembles), which are illustrated in Fig.~\ref{fig:setup}(b). These modes are strongly coupled, such that they are spectrally separate and able to be individually excited~\cite{Kato2018}. In this paper we focus specifically on the observation of the `cavity dark mode' (mode (v) in Fig.~\ref{fig:setup}(b)), and the corresponding observations of remote atom excitation and nonlocal atomic saturation. The two cavities are oscillation nodes of this mode, meaning that the two distant atomic ensembles communicate only via the remote link fiber. We emphasize that this is a truly macroscopic network: the cavities are each of order 1 meter long. This observation of all normal modes of a macroscopically large quantum network, observed simultaneously at two points of the network, lays the foundation for extension to larger networks of multiple atom-cavity systems for quantum information processing purposes.

Let us first consider the system with one atom for each cavity, whose Hamiltonian ($\hbar = 1$) is given by:
\begin{align}
\begin{split}
H &= 
\omega_{\rm c}\left( a_1^\dag a_1 + a_2^\dag a_2 + b^\dag b \right)
+\sum_{i=1,2} v_i \left( a_i^\dag b + b^\dag a_i \right) \\
&+\omega_{\rm a} \left( \sigma_1^+ \sigma_1^- +  \sigma_2^+ \sigma_2^- \right)
+\sum_{i=1,2} g_i \left(a_i^\dag \sigma_i^- + \sigma_i^+ a_i \right),
\end{split}
\end{align}
where we assume that the cavity and fiber modes $(a_1, a_2, b)$ are degenerate with frequency $\omega_{\rm c}$. The coupling rates of cavities 1 and 2 with the fiber are given by 
\begin{eqnarray}
v_{1,2} = \frac{c}{2}\sqrt{\frac{T_{2,3}}{L_{\rm f}L_{1,2}}},
\end{eqnarray}
where $c$ is the speed of light in the fiber and $T_i$, $L_i$, and $L_{\rm f}$ are the transmittance of the mirror $i$, length of the cavity $i$, and length of the connecting fiber, respectively.
The atoms are coupled to their respective cavity modes with strengths $g_1$ and $g_2$.
The eigenstates of the above Hamiltonian are given by superpositions of certain combinations of the atom excitations, and the photons in the two cavities and the fiber. The eigenstates for the first excited states are given by the superpositions of the base states $|{\rm A}_1\rangle = |e,g,0,0,0\rangle$, $|{\rm A}_2\rangle = |g,e,0,0,0\rangle$, $|{\rm C}_1\rangle = |g,g,1,0,0\rangle$, $|{\rm C}_2\rangle = |g,g,0,1,0\rangle$, $|{\rm C}_{\rm f}\rangle = |g,g,0,0,1\rangle$, where $|i_1,i_2,n_1,n_2,n_{\rm f}\rangle$ denotes the state of the total system with atom 1 and 2 in the states $i_1$ and $i_2$; and cavity 1, 2, and the fiber in the Fock states of photon numbers $n_1$, $n_2$, and $n_{\rm f}$. Specifically, for the simple case of $\omega_{\rm c}=\omega_{\rm a}\equiv \omega_0$, $g_1 = g_2 \equiv g$, and $v_1 = v_2 \equiv v$, the eigenstates and eigenenergies are given by:

\begin{flalign*}
{\rm (i)} \quad & |{\rm BS1}\rangle \propto g|{\rm A}_1\rangle + g|{\rm A}_2\rangle + \zeta|{\rm C}_1\rangle &~\nonumber \\ &+ \zeta|{\rm C}_2\rangle + 2v|{\rm F}\rangle, &\omega_0 + \zeta,
\nonumber \\
{\rm (ii)} \quad & |{\rm BS2}\rangle \propto g|{\rm A}_1\rangle + g|{\rm A}_2\rangle  -\zeta|{\rm C}_1\rangle \nonumber \\ &- \zeta|{\rm C}_2\rangle + 2v|{\rm F}\rangle, & \omega_0 - \zeta,
\nonumber \\
{\rm (iii)} \quad & |{\rm FD1}\rangle \propto |{\rm A}_1\rangle - |{\rm A}_2\rangle  +|{\rm C}_1\rangle - |{\rm C}_2\rangle, & \omega_0 + g,
\nonumber \\
{\rm (iv)} \quad & |{\rm FD2}\rangle \propto |{\rm A}_1\rangle - |{\rm A}_2\rangle  -|{\rm C}_1\rangle + |{\rm C}_2\rangle, & \omega_0 - g,
\nonumber \\
{\rm (v)} \quad & |{\rm CD}\rangle \propto v|{\rm A}_1\rangle +v|{\rm A}_2\rangle - g|{\rm F}\rangle, & \omega_0,
\nonumber
\end{flalign*}

\noindent where $\zeta=\sqrt{g^2+2v^2}$ is the symmetric mode resonance shift. The modes (i)-(v) are illustrated in Fig.~\ref{fig:setup}(b). The two states of (i) $|{\rm BS1}\rangle$ and (ii) $|{\rm BS2}\rangle$ are `bright states' and have photon excitations in the two cavities and the fiber. 
In contrast, the other three are `dark states', where photon excitations are absent either from the link fiber or from the two end cavities. Two states --- the `fiber dark states' of (iii) $|{\rm FD1}\rangle$ and (iv) $|{\rm FD2}\rangle$ --- do not exhibit excitation in the central link fiber. 
Of particular interest is the `cavity dark state' (v) $|{\rm CD}\rangle$, which has no photon excitation in the two cavities in which atoms are placed. In other words, it is the state of atoms dressed with the remote photons of the link fiber.  We emphasize that this state exists only when atoms are coherently coupled to both cavities. For the general case with $g_1 \neq g_2$ and $v_1 \neq v_2$, the states of (iii) $|{\rm FD1}\rangle$ and (iv) $|{\rm FD2}\rangle$ are no longer pure `fiber dark' states, although for the parameters discussed in this work the fiber contribution is negligibly small (see Supplementary Material~\cite{SM2019}). The state (v)  $|{\rm CD}\rangle$ remains a pure `cavity dark' state independent of cavity symmetry.

These eigenstates (i) - (v) correspond to the normal modes of the system dynamics in the weak-driving limit~\cite{SM2019}, as illustrated in Fig.~\ref{fig:setup}(b). For a system with ensembles of atoms in the cavities, the linear optical response in the weak-driving limit is identical to the single-atom model, in which the single-atom coupling strengths $g_i$ are replaced by the effective coupling strengths $g_{i, {\rm eff}} = g_{i} \sqrt{N_{i, {\rm eff}}}$, where $N_{i, {\rm eff}}$ is the effective number of atoms in cavity $i$~\cite{Kato2018}.

The setup is similar to that of our previous work~\cite{Kato2018}, where we observed four of the five normal modes: the two symmetric modes (i) and (ii), and the two fiber dark modes (iii) and (iv). In our previous work, we directly excited Cavity 1 and detected the response of Cavity 2. Both of these cavities are nodes of the cavity dark mode (v), meaning that we could not detect this mode in the original work. In this experiment, we introduce a fiber beamsplitter into the link fiber to study this unique mode of oscillation. Specifically, the setup shown in Fig.~\ref{fig:setup}(a) is designed to allow the system to be driven and detected at either the end of the cavity array (ports $A$ and $B$), or through a fiber beamsplitter at the central link fiber (port $C$).  The response of one network node at port $B$ is simultaneously observed with the response of the link fiber at port $C$. This enables experimental probing of all normal modes, and provides simultaneous access to cavity oscillation nodes and antinodes. The weak 1\% outcoupling of the beamsplitter ensures that the normal modes are not excessively broadened by loss.

An experimental run consists of three main steps. Firstly, laser-cooled Cs atoms in the $6^{2}S_{1/2}~F=4$ state are loaded from a magneto-optical trap into a compensated evanescent-field far-off-resonant dipole trap (FORT)~\cite{LeKien2005,Vetsch2010,Lacroute2012,Goban2012}. An optical lattice of 937 nm light and a repulsive 688 nm beam are present in the nanofiber region to form the series of trap sites. Secondly, spectroscopy is performed on the atom-cavity system by sweeping a probe laser, input either at port $A$ or $C$, from -30 to +30 MHz with respect to the atomic and bare-cavity resonances. Thirdly, the atoms are optically pumped into the dark $F=3$ state, and spectroscopy is performed on the effectively empty-cavity condition with input at port $A$. Single-photon counting modules (SPCMs) detect the response at ports $B$ and $C$ for both frequency sweeps.

\begin{figure}[t]
\centering
\includegraphics[width=8.0cm]{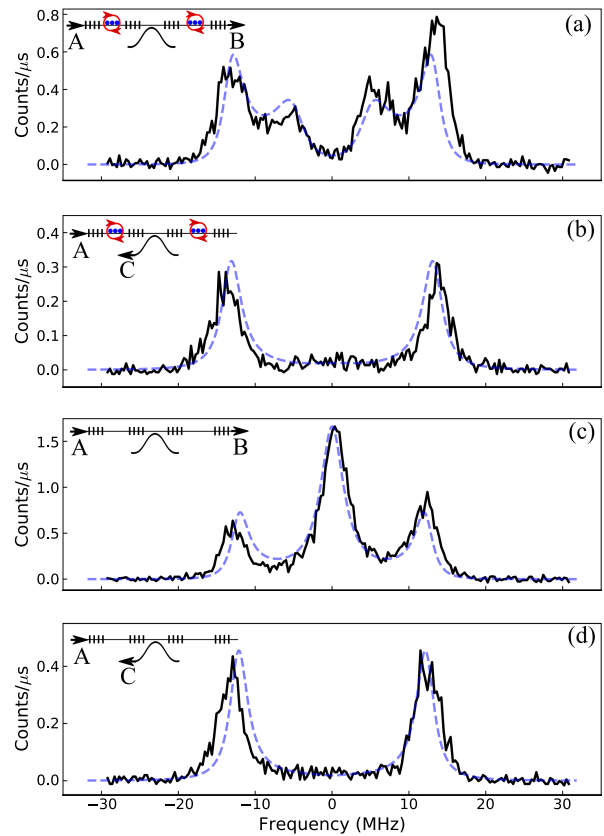}
\caption{Probing the fiber dark mode. (a)-(d) show data for the spectroscopy driving at port $A$ for ($L_1$, $L_f$, $L_2$) = (0.92, 1.40, 1.38) m and FBG reflectances (0.85,0.57,0.72,0.85). Data is overlaid with calculations performed with the single-mode linearized model~\cite{SM2019}. The bare atomic and single-cavity resonances are located at 0 MHz. (a) Atoms are in both cavities and output is detected at port $B$. The fiber dark mode is visible as the doublet at $\pm$5~MHz. The two symmetric bright modes are also observed at $\pm$13.6~MHz. (b) Atoms are in both cavities and output is detected at port $C$. The fiber dark mode is absent, and only the two symmetric bright modes are observed. (c) Empty cavity spectra detected at port $B$. The central peak and the two sideband peaks correspond to the fiber dark mode and the two symmetric bright modes for coupled empty cavities. (d) Empty cavity spectra detected at port $C$. The fiber dark mode is absent. The probe drive-strength at Port $A$ is 250 pW.}
\label{fig:A-driving}
\end{figure}

\begin{figure}[th!]
\centering
\includegraphics[width=8.2cm]{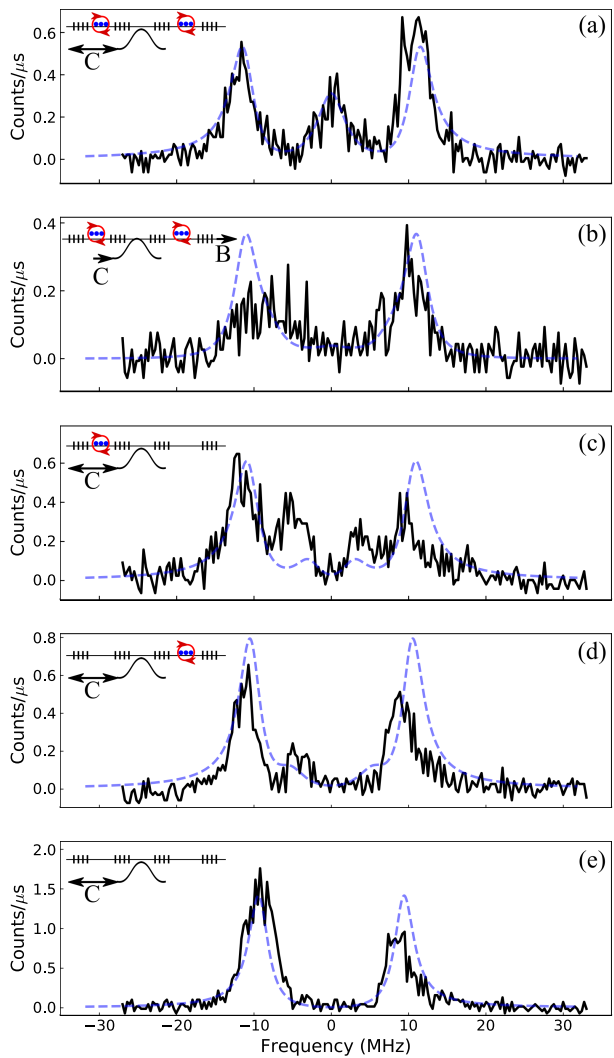}
\caption{Probing the cavity dark mode. (a)-(d) show data for the spectroscopy driving and detecting at port $C$ for ($L_1$, $L_f$, $L_2$) = (0.92, 1.80, 1.38) m and FBG reflectances (0.80,0.65,0.80,0.85).  Dashed lines show theoretical calculations performed with the single-mode linearized model~\cite{SM2019}. (a) Atoms are in both cavities ($C\rightarrow C$ spectroscopy). The cavity dark mode is visible as the central 0 MHz resonance. The two symmetric bright modes are also observed. (b) Atoms are in both cavities ($C\rightarrow B$ spectroscopy). Only the two bright modes are observed. (c) Atoms are in Cavity 1 only ($C\rightarrow C$ spectroscopy). Four normal modes are observed. (d) Atoms are in Cavity 2 only ($C\rightarrow C$ spectroscopy), and four normal modes are observed. (e) Empty cavity spectra ($C\rightarrow C$ spectroscopy), where two normal modes are observed. The input probe power at Port $C$ is 800~pW.}
\label{fig:C-driving}
\end{figure}

We first show the results of driving the input port $A$ in Fig.~\ref{fig:A-driving}, similar to our previous work~\cite{Kato2018}, but this time measuring the two output ports $B$ and $C$ simultaneously. The $A \rightarrow B$ transmissions in Fig.~\ref{fig:A-driving}(a) and (c) reproduce the observation of the `fiber' dark modes and the bright modes in Ref.~\cite{Kato2018}. Furthermore, the naming of the `fiber dark' modes is supported by the suppression of the corresponding peaks in the $A \rightarrow C$ transmissions in Fig.~\ref{fig:A-driving}(b) and (d). Only the bright modes have excitation at both the end cavity and the link fiber, and therefore can be driven and detected in this $A\rightarrow C$ configuration. The data agrees with the theoretical curve of the steady-state solution for the linearized master equation in the weak-driving limit with $(g_{1, {\rm eff}}, g_{2, {\rm eff}}) = (5.0, 5.0)$~MHz~\cite{SM2019}. All theoretical curve amplitudes have been scaled based on the peak empty cavity response.

Next we show the results of driving and detecting the port $C$ in Fig.~\ref{fig:C-driving}.
The central result of this experiment is the observation of the cavity dark mode at the atomic resonance (0 MHz) in Fig.~\ref{fig:C-driving}(a). This mode is absent when driving the port $A$ in Fig.~\ref{fig:A-driving}(a) and (b), due to the direct excitation of cavity photons. Figure~\ref{fig:C-driving}(b) indicates on-resonant suppression of the output at Port B, confirming that the `cavity dark mode' does not support photonic excitations within the cavities. We note that the cavity dark mode signature is only observed in the case where both atomic ensembles are coupled to the cavities. In cases where atoms are coupled only to single cavities (Fig.~\ref{fig:C-driving}(c) and (d)), the $C\rightarrow C$ resonant transmission is suppressed. In these singly-loaded cases, we may interpret the experiment in two ways. Firstly, we can consider the interaction of atoms with the on-resonant empty-cavity fiber dark mode~\cite{Kato2018}. This induces a vacuum Rabi splitting of the fiber dark mode, resulting in the observation of four unique spectroscopic peaks, and on-resonant suppression. Alternatively, one may view the system as the collective oscillation of four oscillators (three cavities and one atom), manifesting as two symmetric and two antisymmetric modes. In the case of four coupled oscillators, all oscillators are antinodes for all modes, resulting in their observation in Fig.~\ref{fig:C-driving}(c) and (d). In all cases, the data is in agreement with the theoretical curve with $(g_{1, {\rm eff}}, g_{2, {\rm eff}}) = (6.0, 7.0)$~MHz~\cite{SM2019}.

We note that the empty cavity responses in Figs.~\ref{fig:A-driving}(c),~\ref{fig:A-driving}(d) and~\ref{fig:C-driving}(e) may be recovered in the model by setting $g=0$, resulting in $|{\rm FD1}\rangle$ and $|{\rm FD2}\rangle$ coalescing to a single fiber dark mode, while $|{\rm CD}\rangle$ vanishes.

\begin{figure}[t]
\centering
\includegraphics[width=8.2cm]{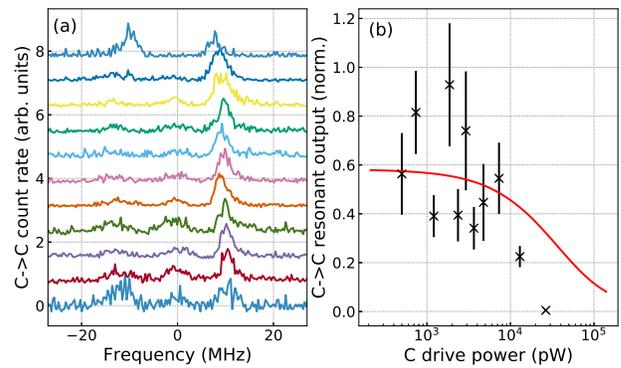}
\caption{Saturation of the dark mode. (a) $C\rightarrow C$ transmission curves for atoms loaded into both cavities, where increasing offset indicates increasing drive power (from 0.50 nW to {27~nW}). (b) The on-resonant transmission is normalized with respect to the average amplitude of the two bright modes. The errorbars are statistical. A theoretical curve is overlaid in red.}
\label{fig:saturation}
\end{figure}

The above observation of the cavity dark mode for $C\rightarrow C$ transmission at zero detuning can be interpreted as \textit{remote} excitation of atoms through the excitation of photons in the link fiber. Although no local photons are excited at the atom locations, we expect this dressed state to saturate at high drive powers. We expect the system response at high drive intensities to tend toward the empty-cavity dual peak spectrum of Fig.~\ref{fig:C-driving}(e), and result in the dark mode signal diminishing with increasing intensity of excitation at the link fiber due to \textit{remote} saturation of atoms.

Figure~\ref{fig:saturation}(a) confirms this hypothesis. The on-resonant peak is clearly resolved at low drive intensities, and is absent at high intensities. We therefore obtain the counter-intuitive result that increasing drive strength \textit{reduces} the on-resonant response of the link fiber cavity. We emphasize that the atoms do not experience \textit{local} intensity on-resonance, because driving at $C$ excites the cavity dark mode in the low intensity limit, and does not excite the fiber dark mode in the high intensity empty-cavity limit. The $C\rightarrow C$ saturation theoretical curve is obtained by solving the coupled semiclassical equations of motion describing the nonlinear dynamics~\cite{SM2019}, and agrees qualitatively with the cavity dark mode amplitudes plotted in Fig.~\ref{fig:saturation}(b). We attribute the enhanced saturation observed in the experiment to the asymmetric drive of the link fiber, which introduces a nonzero light level in Cavity 2 for empty cavities. The saturation values used in the model are derived from a separate experiment driving from $A\rightarrow B$ for atoms in single cavities. From this data, we measure atom numbers of 370 and 250, saturation photon numbers of 40 and 20, and many-atom coupling strengths of 6.0 and 7.0 MHz in Cavities 1 and 2 respectively. 

We note that the asymmetry observed between the low- and high-frequency sides of the spectra in Figs.~\ref{fig:A-driving}~--~\ref{fig:saturation} when atoms are coupled to the cavities arises from the light shift of the off-resonant probe beam~\cite{SM2019}. 

In conclusion, we have observed all five normal modes of a large coupled cavity QED system. In particular, we have observed the cavity dark mode, which is an excitation of atoms dressed with photons in a cavity which does not couple directly to either atomic ensemble. The nonlinear response of this mode shows remote excitation and saturation of atoms without photon excitations at the atom locations. 
We are especially interested in improving this system by overcoming technical challenges related to the simultaneous resonant locking of $N>1$ optical cavities, and the trapping of single atoms in networked Cavity QED systems.


\appendix

\section{Experimental methods}

This is a spectroscopic experiment performed on a system of three optical cavities, coupled to two atomic ensembles. The spectrum depends on the resonant frequencies of the individual cavities with respect to the atoms, and we impose the condition that each of the three cavities is resonant with the $F=4\rightarrow 5$ D2 line of the Cs atomic resonance. The three cavities are not actively locked to the atomic resonance, and the experiment requires both `pre-selection' triggering and `post-selection' on features of the data. The pre-selection is performed by monitoring the $A\rightarrow B$ transmission of an on-resonance probe while the three cavity lengths are modulated at incommensurate frequencies. An experimental run is triggered above a defined transmission threshold, and the cavity-length-modulating piezoelectric transducer voltages are maintained at constant values through the duration of the run (with a total time of 145 ms, for $N_{rpt}=5$ sequence repeats of atom cooling and spectroscopy). The post-selection is performed on the empty cavity data, by ensuring that the empty-cavity $A\rightarrow B$ spectrum is maximum at 0 MHz, and that there is sufficient sideband intensity in the $A\rightarrow C$ data. This ensures that an optimum resonance condition has been met, such that all three cavities are resonant with the atomic transition. 

Following the pre-selection trigger, a typical experimental run begins with excitation at port $C$. Atoms are loaded into both cavities via standard magneto-optical trap (MOT) cooling. The atoms are trapped in a two-color magic wavelength evanescent far-off-resonant dipole trap (FORT), formed from an optical lattice of two counterpropagating 100 $\mu$W 937 nm beams and a single-pass of a 5.2 mW 688 nm beam. The system is probed spectroscopically, with the input probe swept in frequency by 60 MHz across the atomic resonance in 4 ms, which is significantly slower than the timescales of the system dynamics. The system is simultaneously detected at Ports $B$ and $C$. The output light is first filtered through a series of polarization and interferometric filters, to remove unwanted light from FORT beams and from the background. Following the filtering system, the light is detected by single photon counting modules (SPCMs). The detected photon streams are binned in time to experimentally obtain a spectroscopic response curve at both ports $A$ and $C$. The dual detection scheme allows for the different responses of the cavities and the link fiber to be experimentally measured, enabling the study of the fiber and cavity dark modes.

Following the spectroscopy with atoms coupled to cavities, the empty-cavity spectrum is probed. Optical pumping beams, resonant with the D2 $F=4\rightarrow 3$ transition, pump the atomic ensembles into the $F=3$ ground state, to ensure that they are `dark' to the cavity which is resonant to the $F=4\rightarrow 5$ transition. A micro-electromechanical-systems (MEMS) optical switch swaps the input port to $A$, and the $A\rightarrow B$ and $A\rightarrow C$ empty cavity signals are probed. Optical cooling then commences with the same atomic ensemble to replenish the optical dipole trap, which has a lifetime of $\approx$ 10 ms. The sequence repeats $N_{rpt}=5$ times, using the same cavity condition. A new MOT is then loaded, the cavity length modulation restarts, and the process repeats when the cavity trigger threshold is met.

\begin{widetext}

\section{Single-mode quantum model}

\begin{figure}[b]
\centerline{
\includegraphics[width=0.8\textwidth]{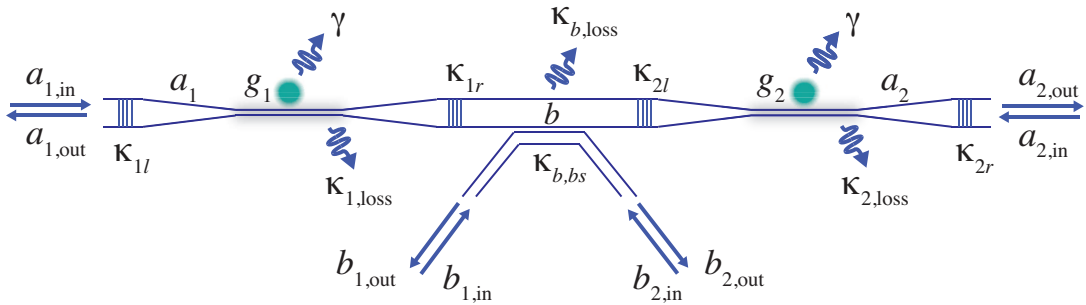}
}
\caption{Schematic of the coupled-cavities system (not to scale).}\label{fig:config}
\end{figure}

As a simple, quantum model of our system (Fig.~\ref{fig:config}), we consider single modes for the fields in the cavities (annihilation operators $a_1$ and $a_2$) and in the connecting fiber ($b$), and single, two-level atoms in each cavity (raising and lowering operators $\sigma_1^\pm$ and $\sigma_2^\pm$). Then, a master equation for the density operator $\rho$ of the total system, in a frame rotating at the probe laser frequency $\omega_{\rm p}$ (assumed the same for all driving fields in the model), takes the form
\begin{align}\label{eq:ME1}
\dot{\rho} &= -i[H,\rho ] + \kappa_1' {\cal D}[a_1]\rho + \kappa_2' {\cal D}[a_2]\rho + \kappa_b' {\cal D}[b]\rho 
+ \frac{\gamma_\parallel}{2} \left( {\cal D}[\sigma_1^-]\rho + {\cal D}[\sigma_2^-]\rho \right) 
\nonumber
\\
& ~~~~~~~ + \gamma_{\rm las} \left( {\cal D}[a_1^\dag a_1]\rho + {\cal D}[a_2^\dag a_2]\rho + {\cal D}[b^\dag b]\rho 
+ {\cal D}[\sigma_{z,1}]\rho + {\cal D}[\sigma_{z,2}]\rho \right) .
\end{align}
where ${\cal D}[O]\rho =2O\rho O^\dag - O^\dag O\rho -\rho O^\dag O$, and we set $\hbar=1$. The Hamiltonian can be written as $H=H_{\rm sys}+H_{\rm drive}$, with
\begin{align} \label{eq:Hsys}
H_{\rm sys} &= \Delta_1 a_1^\dag a_1 + \Delta_2 a_2^\dag a_2 + \Delta_b b^\dag b + \left( v_1 a_1^\dag b + v_1^\ast b^\dag a_1 \right) + \left( v_2 a_2^\dag b + v_2^\ast b^\dag a_2 \right) \nonumber
\\
& ~~~ + \Dela \left( \sigma_1^+\sigma_1^- + \sigma_2^+\sigma_2^- \right) +  \left( g_1 a_1^\dag \sigma_1^- + g_1^\ast \sigma_1^+ a_1 \right) + \left( g_2 a_2^\dag \sigma_2^- + g_2^\ast \sigma_2^+ a_2 \right) ,
\end{align}
and
\begin{align}
H_{\rm drive} &= \left( {\cal E}_1^\ast a_1 + {\cal E}_1 a_1^\dag \right) +  \left( {\cal E}_2^\ast a_2 + {\cal E}_2 a_2^\dag \right) +  \left( {\cal E}_b^\ast b + {\cal E}_b b^\dag \right) .
\end{align}
The detunings are given by $\Delta_1=\omega_{a_1}-\omega_{\rm p}$, $\Delta_2=\omega_{a_2}-\omega_{\rm p}$, $\Delta_b=\omega_b-\omega_{\rm p}$, and $\Delta_{\rm a}=\omega_{\rm a}-\omega_{\rm p}$. The Hamiltonian $H_{\rm drive}$ describes probe laser fields of amplitudes ${\cal E}_1$, ${\cal E}_2$, and ${\cal E}_b$ driving the cavity and fiber modes through the input channels $a_{1,{\rm in}}$, $a_{2,{\rm in}}$, and $b_{1,{\rm in}}$, respectively. The atoms couple with strengths $g_{1,2}$ to their respective cavity modes, while the coupling rates between the cavity modes (of lengths $L_{1,2}$) and the fiber mode (of length $L_{\rm f}$) are given by
\begin{align}
v_1 = \sqrt{\frac{\kappa_{1r}}{\pi} \omega_{\rm FSR,f}} \equiv \frac{c}{2}\sqrt{\frac{T_2}{L_1L_{\rm f}}} , ~~~~ v_2 = \sqrt{\frac{\kappa_{2l}}{\pi} \omega_{\rm FSR,f}} \equiv \frac{c}{2}\sqrt{\frac{T_3}{L_2L_{\rm f}}} .
\end{align}
Here, $\omega_{\rm FSR,f}=\pi c/L_{\rm f}$ is the free spectral range of the coupling fiber mode, where $c$ is the speed of light in the fiber, and $\kappa_{1r}=cT_2/(4L_1)$ and $\kappa_{2l}=cT_3/(4L_2)$ correspond to the decay rates of the respective cavity fields through mirrors 2 and 3 in the case that the outputs from these mirrors couple to a continuum of modes (e.g., in the limit that $L_{\rm f}\rightarrow\infty$).

The remaining terms in the master equation describe losses and dephasing effects in the system. The cavity and fiber fields decay with rates
\begin{align}
\kappa_1' = \kappa_{1l} + \kappa_{\rm 1,loss} , ~~~ \kappa_2' = \kappa_{2r} + \kappa_{\rm 2,loss} , ~~~ \kappa_b' = \kappa_{b,bs} + \kappa_{b,{\rm loss}} ,
\end{align}
where $\kappa_{1l}=cT_1/(4L_1)$, $\kappa_{2r}=cT_4/(4L_2)$, and, for a 99/1 beamsplitter,
\begin{align}
\kappa_{b,bs} = - \frac{1}{2} \frac{c}{L_{\rm f}} \ln (0.99) .   
\end{align}
The intrinsic loss rates are determined from the (intensity) transmission coefficients of the fiber segments that support the various modes as
\begin{align}
\kappa_{\rm 1,loss} = - \frac{1}{2} \frac{c}{L_1} \ln (1-\alpha_1) , ~~~ \kappa_{\rm 2,loss} = - \frac{1}{2} \frac{c}{L_2} \ln (1-\alpha_2) , ~~~ \kappa_{b,{\rm loss}} = - \frac{1}{2} \frac{c}{L_{\rm f}} \ln (1-\alpha_{\rm f}) ,
\end{align}
where $\alpha_1$, $\alpha_2$, and $\alpha_{\rm f}$ are single-pass losses for the segments in cavity 1, cavity 2, and the connecting fiber, respectively. Finally, the term proportional to $\gamma_{\rm las}$ -- the laser linewidth (HWHM) -- is included so as to incorporate the effect of laser frequency fluctuations, which appears as phase damping of the field and atomic amplitudes. The atoms decay into free space with rate $\gamma_\parallel$.

\subsection{General Normal Modes}

Consider the system Hamiltonian (\ref{eq:Hsys}) for the case in which the cavity and fiber modes are resonant with the atomic transition frequency, and $\Delta_1=\Delta_2=\Delta_b=\Delta_{\rm a}=0$. The resulting Hamiltonian may be diagonalized to give the normal mode operators
\begin{align}
\label{eq:darkOp}
\Dop &= \frac{1}{GZ}\lk g_2v_1\sigop_1+g_1v_2\sigop_2-g_1g_2\bop\rk , &  \tilde{\omega}_D &= 0, \\
\Apmop & = \frac{1}{2\delta G}\lsz g_1V_-\sigop_1-g_2V_+\sigop_2  \pm G\lk V_-\aop_1-V_+\aop_2\rk - \frac{N^2}{V_+}v_2\bop\rsz , & \tilde{\omega}_{A,\pm}&=\pm G,\\
\Spmop & = \frac{1}{2\delta Z}\lsz g_1V_+\sigop_1+g_2V_-\sigop_2  \pm Z\lk V_+\aop_1+V_-\aop_2\rk + \frac{W^2}{V_-}v_2\bop\rsz , &\tilde{\omega}_{S,\pm}&=\pm Z ,
\end{align}
where
\begin{align*}
G &= \sqrt{\bar{g}^2+\bar{v}^2-\delta^2} , & Z &= \sqrt{\bar{g}^2+\bar{v}^2+\delta^2} , \\
N &= \sqrt{\tilde{g}^2-\bar{v}^2+\delta^2} , & W &= \sqrt{-\tilde{g}^2+\bar{v}^2+\delta^2} , \\
V_\pm &= \sqrt{\delta^2\pm\lk\tilde{g}^2+\tilde{v}^2\rk} , & \delta^2 &= \sqrt{\lk\tilde{g}^2+\tilde{v}^2\rk^2+v_1^2v_2^2} , \\
\tilde{g}^2 &= \frac{g_1^2-g_2^2}{2} , & \bar{g}^2 &= \frac{g_1^2+g_2^2}{2} , \\
\tilde{v}^2 &= \frac{v_1^2-v_2^2}{2} , & \bar{v}^2 &= \frac{v_1^2+v_2^2}{2} .
\end{align*}

\noindent It is clear from (\ref{eq:darkOp}) that the cavity dark mode $\tilde{D}$ has zero contribution from the cavity modes $\hat{a}_{1,2}$. The fiber dark modes $\tilde{A}_{\pm}$ have a minor contribution from the fiber mode $\hat{b}$, which vanishes in the case of symmetric cavities ($g_1=g_2=g$, $v_1=v_2=v$).
\begin{align}
\label{eq:sym_lim}
G&\rightarrow g , & Z&\rightarrow\zeta=\sqrt{g^2+2v^2} , & N&\rightarrow 0 , & W&\rightarrow \sqrt{2}v , & V_\pm&\rightarrow v , &\delta&\rightarrow v ,
\end{align}
and thus we obtain back the expression in the main text,
\begin{align*}
\Dop&\rightarrow\hat{D} = \frac{1}{\zeta}\lsz v\lk\sigop_1+\sigop_2\rk - g\bop\rsz , &  \omega_D &= 0 , \\
\Apmop&\rightarrow\hat{A}_\pm = \frac{1}{2}\lsz \lk\sigop_1-\sigop_2\rk \pm \lk\aop_1-\aop_2\rk\rsz , &  \omega_{A,\pm} &= \pm g , \\
\Spmop &\rightarrow\hat{S}_\pm = \frac{1}{2\zeta}\lsz g\lk\sigop_1+\sigop_2\rk  \pm \zeta\lk \aop_1+\aop_2\rk + 2v\bop\rsz , & \omega_{S,\pm}&=\pm \zeta .
\end{align*}

For our experimental parameters, the fiber dark modes exhibit a fiber excitation less than $10^{-5}$ of the atoms or cavity, i.e. $\left|\langle 0|\hat{b}\tilde{A}_{\pm}^\dag |0\rangle\right|^2/\left|\langle 0|\hat{X}\tilde{A}_{\pm}^\dag |0\rangle\right|^2<10^{-5}$, where $\hat{X}=\hat{\sigma}_{1,2}^-,\hat{a}_{1,2}$. The data of Fig.~2(a) and (b) of the main manuscript highlights the absence of photonic excitation in the fiber when the fiber dark mode is excited.


We note that the Hamiltonian $H$ may be expressed in terms of the normal mode operators. This is a lengthy expression which we will not express in full here. However, in the case that the mode splittings $G$ and $Z$ are significantly larger than the decay rates, a rotating wave approximation may be made. In this case, we may focus only on the contribution of the cavity dark mode $\tilde{D}$. In the case where the cavity and fiber modes are resonant with the atomic transition, $\Delta_{1,2}=\Delta_b=\Delta_{\rm a}=\Delta$, and for $\Delta\ll G,Z$, the Hamiltonian reduces to the simple form
\begin{equation}
\tilde{H} = \Delta\tilde{D}^\dag\tilde{D}.
\end{equation}
This form does not depend on a symmetric cavity condition.

\subsection{Linearised model}

If we assume weak driving and, hence, weak excitation of the atoms, then we may derive the following linear equations of motion for the field and atomic amplitudes,
\begin{align}\label{eq:LEOM}
\dot{\braket{a_1}} &= -(\kappa_1+i\Delta_1)\braket{a_1} - iv_1\braket{b} - i g_1\braket{\sigma_1^-} - i{\cal E}_1 ,
\\
\dot{\braket{a_2}} &= -(\kappa_2+i\Delta_2)\braket{a_2} - iv_2\braket{b} - i g_2\braket{\sigma_2^-} - i{\cal E}_2 ,
\\
\dot{\braket{b}} &= -(\kappa_b+i\Delta_b)\braket{b} - iv_1^\ast \braket{a_1} - iv_2^\ast \braket{a_2} - i{\cal E}_b ,
\\
\dot{\braket{\sigma_1^-}} &= -(\gamp+i\Dela)\braket{\sigma_1^-} - i g_1^\ast \braket{a_1}  ,
\\
\dot{\braket{\sigma_2^-}} &= -(\gamp+i\Dela)\braket{\sigma_2^-} - i g_2^\ast \braket{a_2} ,
\end{align}
where $\kappa_1=\kappa_1'+\gamma_{\rm las}$, $\kappa_2=\kappa_2'+\gamma_{\rm las}$, $\kappa_b=\kappa_b'+\gamma_{\rm las}$, and $\gamp=\gamma/2+\gamma_{\rm las}$.
Setting the time derivatives to zero, we find the general steady state solution for the amplitude of cavity 2 as
\begin{align}\label{eq:a2ss}
\braket{a_2}_{\rm ss} = \frac{A}{B} \, ,
\end{align}
where
\begin{align}\label{eq:A}
A =& i{\cal E}_2 + {\cal E}_b \left( \frac{v_2}{\kappa_b+i\Delta_b}\right) \dfrac{\kappa_1+i\Delta_1+\dfrac{|g_1|^2}{\gamp+i\Dela}}{\kappa_1+i\Delta_1+\dfrac{|g_1|^2}{\gamp+i\Dela}+\dfrac{|v_1|^2}{\kappa_b+i\Delta_b}} \nonumber
\\ 
& - i{\cal E}_1 \left( \frac{v_2}{\kappa_b+i\Delta_b}\right) \frac{v_1^\ast}{\kappa_1+i\Delta_1+\dfrac{|g_1|^2}{\gamp+i\Dela}+\dfrac{|v_1|^2}{\kappa_b+i\Delta_b}} ,
\end{align}
and
\begin{align}\label{eq:B}
B = -(\kappa_2+i\Delta_2) - \frac{|v_2|^2}{\kappa_b+i\Delta_b} - \frac{|g_2|^2}{\gamp+i\Dela} + \frac{|v_1v_2|^2}{(\kappa_b+i\Delta_b)^2}\, \frac{1}{\kappa_1+i\Delta_1+\dfrac{|g_1|^2}{\gamp+i\Dela}+\dfrac{|v_1|^2}{\kappa_b+i\Delta_b}} .
\end{align}
Solutions for the steady state amplitudes of cavity 2 and fiber mode $b$ then follow from
\begin{align}\label{eq:a1ss}
\braket{a_1}_{\rm ss} = -\frac{i{\cal E}_1+{\cal E}_b \dfrac{v_1}{\kappa_b+i\Delta_b}+ \dfrac{v_1v_2^\ast}{\kappa_b+i\Delta_b}\,\braket{a_2}_{\rm ss}}{\kappa_1+i\Delta_1+\dfrac{|g_1|^2}{\gamp+i\Dela}+\dfrac{|v_1|^2}{\kappa_b+i\Delta_b}} ,
\end{align}
and
\begin{align}\label{eq:bss}
\braket{b}_{\rm ss} = -\frac{i{\cal E}_b}{\kappa_b+i\Delta_b} - \frac{iv_1^\ast}{\kappa_b+i\Delta_b}\,\braket{a_1}_{\rm ss} - \frac{iv_2^\ast}{\kappa_b+i\Delta_b}\,\braket{a_2}_{\rm ss}  .
\end{align}

To obtain the output photon fluxes, we require the input-output relations,
\begin{align}\label{eq:inputoutput}
a_{\rm 1,out} = a_{\rm 1,in} + \sqrt{2\kappa_{1l}}\, a_1 , ~~~
a_{\rm 2,out} = a_{\rm 2,in} + \sqrt{2\kappa_{2r}}\, a_2 , ~~~
b_{\rm 1,out} = b_{\rm 1,in} + \sqrt{2\kappa_{b,bs}}\, b ,
\end{align}
where
\begin{align}
\braket{a_{\rm 1,in}} = \frac{i{\cal E}_1}{\sqrt{2\kappa_{1l}}} \, , ~~~
\braket{a_{\rm 2,in}} = \frac{i{\cal E}_2}{\sqrt{2\kappa_{2r}}} \, , ~~~
\braket{b_{\rm 1,in}} = \frac{i{\cal E}_b}{\sqrt{2\kappa_{b,bs}}} \, .
\end{align}
In the linear approximation, the output photon fluxes are given by $|\braket{a_{\rm 1,out}}|^2$, $|\braket{a_{\rm 2,out}}|^2$, and $|\braket{b_{\rm 1,out}}|^2$.

\subsection{Saturation}
For driving from $A\rightarrow B$ with atoms only in Cavity 1 or only in Cavity 2, the theoretical analysis of saturation in the system is outlined in detail in \cite{Kato19}, where careful attention is paid to the spatial dependence of the cavity modes and the atomic density. From comparison of this analysis with the experimental data for $A\rightarrow B$ driving, we are able to deduce values for the saturation photon numbers,
\begin{align}
n_{\rm 1,sat} = \frac{\gamp\gamll}{4g_{1,(0)}^2} , ~~~~
n_{\rm 2,sat} = \frac{\gamp\gamll}{4g_{2,(0)}^2} ,
\end{align}
where $g_{l,(0)}$ ($l=1,2$) is the maximum single-atom coupling strength for an atom located at a potential minimum of the dipole trap in cavity $l$, and for the effective atom number $N_{l,{\rm eff}}$ in each cavity. In particular, we find $\{ n_{\rm 1,sat},N_{1,{\rm eff}}, n_{\rm 2,sat},N_{2,{\rm eff}} \} \simeq \{ 40,370,20,250\}$.

For driving from $C\rightarrow C$ the nonlinear semiclassical equations of motion for the field and atomic amplitudes and atomic inversion are slightly modified from those used in \cite{Kato19}. In particular, after eliminating the atomic variables, the coupled, steady state equations for the amplitudes of the cavity and connecting-fiber modes take the form
\begin{align}
0 &= -\braket{a_1} \left\{ \left( \kappa_1 +i\Delta_1 \right) + \left(\gamp - i\Dela \right) \sum_{j_1} \frac{g_{1j_1}^2}{\gamp^2+\Dela^2+4\dfrac{\gamp}{\gamma_\parallel}g_{1j_1}^2\left|\braket{a_1}\right|^2} \right\} - iv_1\braket{b} ,
\\
0 &= -\braket{a_2} \left\{ \left( \kappa_2 +i\Delta_2 \right) + \left(\gamp - i\Dela \right) \sum_{j_2} \frac{g_{2j_2}^2}{\gamp^2+\Dela^2+4\dfrac{\gamp}{\gamma_\parallel}g_{2j_2}^2\left|\braket{a_2}\right|^2} \right\} - iv_2\braket{b} ,
\\
i\Eb &= -\left( \kappa_b+i\Delta_b \right) \braket{b} - iv_1\braket{a_1} - iv_2\braket{a_2} ,
\end{align}
where $g_{lj_l}$ is the coupling strength of atom $j_l$ to cavity $l$ ($l=1,2$). We assume for simplicity that the parameters $\{ g_{1j_1},g_{2j_2}\}$, $\{ v_1,v_2\}$, and $\Eb$ are all real. 

Defining normalized amplitudes
\begin{align}
X_1 = \frac{\braket{a_1}}{\sqrt{n_{\rm 1,sat}}} , ~~~~ 
X_2 = \frac{\braket{a_2}}{\sqrt{n_{\rm 2,sat}}} , ~~~~
X_b = \frac{\braket{b}}{(n_{\rm 1,sat}n_{\rm 2,sat})^{1/4}} , ~~~~
y_b = \frac{\Eb /\kappa_b}{(n_{\rm 1,sat}n_{\rm 2,sat})^{1/4}} ,
\end{align}
and assuming a simple, standing-wave profile for the cavity mode, together with a uniform atomic density along this mode (as a result of incommensurate cavity mode and dipole trap wavelengths), it is possible to approximate the summations in the above equations with integrals, which can be evaluated analytically to give the equations
\begin{align}
0 & = X_1 \left\{ \vphantom{\frac{\left( A\right)^2}{\sqrt{A^2}}} \left( 1 + i\frac{\Delta_1}{\kappa_1}\right) \right. \nonumber
\\
& \left. + \left( 1 - i\frac{\Dela}{\gamp} \right) \frac{2C_1}{1+A} \frac{1}{|X_1|^2} \left[ 1 - \frac{1+\left( \Dela /\gamp \right)^2}{\sqrt{\left( 1+ \left( \Dela /\gamp \right)^2 + A|X_1|^2\right) \left( 1+ \left( \Dela /\gamp\right)^2 + |X_1|^2\right)}} \right] \right\} \nonumber
\\
& \hspace{15mm} + i \frac{v_1}{\kappa_1} \left( \frac{n_{\rm 2,sat}}{n_{\rm 1,sat}} \right)^{1/4} X_b ,
\\
0 & = X_2 \left\{ \vphantom{\frac{\left( A\right)^2}{\sqrt{A^2}}} \left( 1 + i\frac{\Delta_2}{\kappa_2}\right) \right. \nonumber
\\
& \left. + \left( 1 - i\frac{\Dela}{\gamp} \right) \frac{2C_2}{1+A} \frac{1}{|X_2|^2} \left[ 1 - \frac{1+\left( \Dela /\gamp \right)^2}{\sqrt{\left( 1+ \left( \Dela /\gamp \right)^2 + A|X_2|^2\right) \left( 1+ \left( \Dela /\gamp\right)^2 + |X_2|^2\right)}} \right] \right\} \nonumber
\\
& \hspace{15mm} + i \frac{v_2}{\kappa_2} \left( \frac{n_{\rm 1,sat}}{n_{\rm 2,sat}} \right)^{1/4} X_b ,
\\
-iy_b & = \left( 1 + i\frac{\Delta_b}{\kappa_b} \right) X_b + i \frac{v_1}{\kappa_b} \left( \frac{n_{\rm 1,sat}}{n_{\rm 2,sat}} \right)^{1/4} X_1 + i \frac{v_2}{\kappa_b} \left( \frac{n_{\rm 2,sat}}{n_{\rm 1,sat}} \right)^{1/4} X_2  ,
\end{align}
where
\begin{align}
C_1 = N_{\rm 1,eff} \frac{g_{1,(0)}^2}{\kappa_1\gamp} , ~~~~
C_2 = N_{\rm 2,eff} \frac{g_{2,(0)}^2}{\kappa_2\gamp} ,
\end{align}
and $A=0.17$ is a geometric factor related to the cavity mode geometry \cite{Kato19}. 

For a given driving strength, quantified by $y_b$, these equations can be solved numerically to give the scaled amplitudes $X_1$, $X_2$, and $X_b$, and thereby the transmission spectrum through each of the output channels. The input power for driving through $C$ is related to the parameters of the above model by 
\begin{align}
P_{{\rm in},C} = \frac{\Eb^2}{2\kappa_{b,bs}} \hbar\omega_{\rm p} = \frac{\Eb^2}{2\kappa_{b,bs}} \frac{2\pi\hbar c}{\lambda} = \frac{\kappa_b^2}{2\kappa_{b,bs}} \frac{2\pi\hbar c}{\lambda} \sqrt{n_{\rm 1,sat}n_{\rm 2,sat}}\, y_b^2 ,
\end{align}
and the normalized transmission on resonance is plotted against this power to give the saturation curve.

\end{widetext}

\section{Parameters}

The cavity parameters for the experiments are listed in Table~\ref{tab:params}. Energy is lost to the environment via spontaneous emission ($\gamma_{||}$), intrinsic loss and cavity outcoupling. The intrinsic intracavity loss arises from fusion splice losses and propagation loss at the tapered nanofiber. The outcoupling occurs at mirrors 1, 4, and the central beamsplitter. We also include a term $\gamma_{las}$ to account for the finite probe linewidth.




\begin{table}[ht]
\caption{List of parameters for modeling the experiment. The two different cavity conditions of the experiments in Fig. 2 and Fig. 3 of the main text are tabulated.}
\begin{tabular}{|c|c|c|}
\hline
\textbf{Parameter} & \textbf{Fig 2 data (2$\pi\cdot$MHz)} & \textbf{Fig 3 data (2$\pi\cdot$MHz)} \\ \hline
$\kappa_{1,loss}$  & 0.36                                 & 0.36                                 \\ \hline
$\kappa_{2,loss}$  & 0.24                                 & 0.24                                 \\ \hline
$\kappa_{b,loss}$  & 0.24                                 & 0.18                                 \\ \hline
$\kappa_{b,bs}$    & 0.12                                 & 0.091                                \\ \hline
$\kappa_{1l}$      & 1.33                                 & 1.78                                 \\ \hline
$\kappa_{1r}$      & 3.82                                 & 3.11                                 \\ \hline
$\kappa_{2l}$      & 1.66                                 & 1.18                                 \\ \hline
$\kappa_{2r}$      & 0.89                                 & 0.89                                 \\ \hline
$v_1$              & 9.45                                 & 7.52                                 \\ \hline
$v_2$              & 6.23                                 & 4.64                                 \\ \hline
$\gamma_{||}$      & 5.2                                  & 5.2                                 \\ \hline
$\gamma_{las}$     & 0.36                                 & 0.36                                 \\ \hline
\end{tabular}
\label{tab:params}
\end{table}

As Table \ref{tab:params} shows, the reflectivities of the mirrors are in general significantly less than $1$. These differences represent themselves in the single-mode model as a reduced coupling $v_i$ between the cavities and the fibre, when only one mode is considered. The standing-wave picture in the low reflectance case starts to break down, which means that in the linear regime a travelling-wave description using transfer matrices can more adequately describe the behavior. Therefore, although the experimental data fit the current model with these parameters well, significantly improved agreement was obtained by scaling the coupling strengths $v_i$ with a factor of 1.075 for Fig. 2 and 1.055 for Fig. 3 in the main text. A more detailed justification based on the transfer matrix method can be found in \cite{TMpaper}.

We scaled all the theoretical curves to match the maximum of the experimental data. We note that in Fig. 3, in the case of driving and detecting through port C, the spectrum showed a significant asymmetry (also explained in the next section). Therefore, we applied an extra scaling factor of 0.8 to the theoretical model.

\section{Asymmetry for red- and blue-detuned probe}

We consistently observe an asymmetry in the detected spectra. The modes $\tilde{S}_-$ and $\tilde{A}_-$ have reproducibly larger central frequency detunings than the modes $\tilde{S}_+$ and $\tilde{A}_+$. In addition, the amplitude of the red-detuned mode is generally lower than the blue-detuned mode. This can be explained by considering the light shift of the probe beam on the atoms.

\begin{figure}[ht]
\centering
\includegraphics[width=0.5\textwidth]{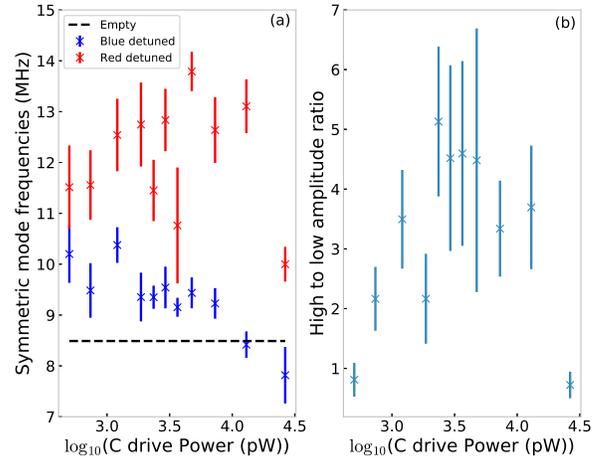}
\caption{Effect of probe light shift on the symmetric modes. (a) Absolute values of the upper and lower mode frequencies (blue- and red-detuned probe respectively) are plotted as a function of drive power. The average value of the empty cavity symmetric mode frequency is shown as a dotted line for reference. (b) The ratio of the peak photon count of the blue-detuned mode $\tilde{S}_+$ to the red-detuned mode $\tilde{S}_-$ as a function of drive power. The data in this figure is measured from the experiment of Fig.~4 of the main manuscript.}
\label{fig:ampratio}
\end{figure}

The light shift alters the equilibrium position of the FORT traps, which modifies the atom--cavity coupling rate $g$. A red-detuned probe attracts the atoms closer to the nanofiber leading to an enhanced $g$, while a blue-detuned probe repels the atoms and reduces $g$. Increased $g$ results in a greater central frequency detuning $\sqrt{g^2+2v^2}$, and also reduces the peak amplitude. Figure~\ref{fig:ampratio} supports this assessment. 

In Fig.~\ref{fig:ampratio}(a), the red-detuned mode's central frequency is observed to increase until the atoms saturate, returning near to the empty cavity frequency for strong driving. In contrast, the blue-detuned central frequency monotonically decreases and approaches the empty cavity mode frequency as the atoms saturate. Similarly, the peak in the amplitude ratio of Fig.~\ref{fig:ampratio}(b) is explained by the increased value of $g$ at weak drive powers. At strong drive powers, saturation causes the spectrum to return to the symmetric empty cavity spectrum. Note that an on-resonant probe beam exciting the cavity dark mode $\tilde{D}$ does not induce a light shift because the cavities are dark, and the light shift from a resonant probe beam is zero.


\section*{Acknowledgments}
The authors acknowledge the support of JST CREST Grant Number JPMJCR1771, JSPS KAKENHI Grant Numbers 16H01055 and 18H04293, and JST PRESTO Grant Number JPMJPR1662, Japan, and Institute for Advanced Theoretical and Experimental Physics, Waseda University. S.P. and N.N. thank the group of T.A. for the hospitality.

\end{document}